\documentclass[aps,prl,twocolumn,showpacs,amsmath,amssymb]{revtex4-1}
\usepackage[dvips]{graphicx}

\newcommand{\be}{\begin{equation}}
\newcommand{\ee}{\end{equation}}

\newcommand{\bea}{\begin{eqnarray}}
\newcommand{\eea}{\end{eqnarray}}
\newcommand{\bd}{\begin{displaymath}}
\newcommand{\ed}{\end{displaymath}}
\newcommand{\ba}{\begin{array}}
\newcommand{\ea}{\end{array}}
\newcommand{\bi}{\begin{itemize}}
\newcommand{\ei}{\end{itemize}}
\newcommand{\bc}{\begin{center}}
\newcommand{\ec}{\end{center}}
\newcommand{\bfl}{\begin{flushleft}}
\newcommand{\efl}{\end{flushleft}}
\newcommand{\bfr}{\begin{flushright}}
\newcommand{\efr}{\end{flushright}}

\def\6{\partial}

\def\={\!\!\!&=&\!\!\!}
\def\+{\!\!\!&&\!\!\!+~}
\def\-{\!\!\!&&\!\!\!-~}

\begin{document}

\title[]{Magnetic resonance from the interplay of frustration and superconductivity}
%
 \author {J. Knolle$^{1}$, I. Eremin$^{2}$, J. Schmalian$^{3}$, and R. Moessner$^{1}$}
 \affiliation{
 $^{1}$Max Planck Institute for the Physics of Complex Systems, D-01187 Dresden, Germany}
 \affiliation {$^2$Institut f\"ur Theoretische Physik III, Ruhr-Universit\"at Bochum, D-44801 Bochum, Germany}
 \affiliation {$^3$Karlsruhe Institute of Technology, Institute for Theory of Condensed Matter, D-76131 Karlsruhe,
Germany}

\begin{abstract}
Motivated by the  iron-based superconductors, we develop a self-consistent electronic theory for the itinerant spin excitations in the regime of coexistence of the antiferromagnetic stripe order with wavevector ${\bf Q}_{1} = (\pi,0)$ and $s^{+-}$ superconductivity. The onset of superconductivity leads to the appearance of a {\em magnetic} resonance near the wavevector ${\bf Q}_{2} = (0,\pi)$ where magnetic order is absent. This resonance is isotropic in spin space, unlike the excitations near ${\bf Q}_{1}$ where the magnetic Goldstone mode resides. We discuss several features which can be observed experimentally.
 \end{abstract}

\date{\today}


\maketitle

The idea of emergent symmetry in strongly correlated
electron systems plays a prominent role in various novel superconductors such as high-T$_c$ cuprates, heavy fermions, and organic compounds. The recent discovery of superconductivity in vicinity  of the antiferromagnetism in iron-based systems\cite{kamihara} has opened up a new perspective in studying electronic instabilities and their symmetry properties. For example, it was argued initially that the close proximity of the antiferromagnetism (AF) and superconductivity (SC) in the iron-based superconductors could be understood within an emergent SO(6) symmetry of the underlying electronic Hamiltonian\cite{podolsky}, a result supported also by a renormalization group study.\cite{Chubukov2008} This symmetry is closely connected to the so-called $s^{+-}$ pairing state of the SC order parameter. In addition, the frustrated AF state with ordering wavevector ${\bf Q}_1 = (\pi, 0)$ or ${\bf Q}_2 = (0,\pi)$ is related to an SU(2) $\times$ Z$_2$ symmetry where the Ising degree of freedom (Z$_2$) refers to the $(\pi,0)$ and $(0,\pi)$ ordering vectors. This is analogous to behavior found in localized magnetic systems, most simply in the frustrated $J_1-J_2$ square lattice model.\cite{chandra}


As such enhanced symmetries are generically not protected, fine details of the electronic structure lift this degeneracy, but only on a scale well below the Fermi energy (about 0.1 eV), and in a way which is ``non-universal``. Even only the selection of the AF stripe order out of the degenerate manifold of the magnetic structures associated with two magnetic order parameters at wave vectors ${\bf Q}_1$ and ${\bf Q}_2$ and connected by the symmetry operations of the $O(6)$ group represents an interesting problem. There, finite ellipticity and interactions between electron pockets remove the degeneracy and favor the observed stripe AF state.\cite{eremin2010} Nevertheless, the competing states in the manifold remain nearby in energy\cite{Knolle2010,yaresko}, resulting in collective excitations at low energy as a remnant of the enhanced symmetry.

AF long range order naturally gives rise to a Goldstone mode. SC of the $s^{+-}$-type leads to a magnetic resonance mode instead.\cite{korshunov,maier} It is an open question how these two rather distinct magnetic collective modes interfere in the case of coexisting AF + SC orders. Of course, the existence of a gapless Goldstone mode (assuming continuous spin symmetry and singlet SC) is guaranteed based on general symmetry principles. It seems plausible to speculate that that the resonance mode is simply absorbed by the Goldstone mode. Here we show that the enhanced Z$_2$-symmetry and the existence of two potential ordering vectors does allow for a strong resonance mode at ${\bf Q_2}=(0,\pi)$ in the coexistence phase if the magnetic order corresponds to ${\bf Q_1}=(\pi,0)$ and vice versa.
This effect is peculiar to itinerant magnetic systems with nearly degenerate ground states and demonstrates that the collective behavior of various excitations may uncover the hidden symmetry of the electronic system.

\begin{figure}[h]
\centering
\includegraphics[width=1.0\linewidth]{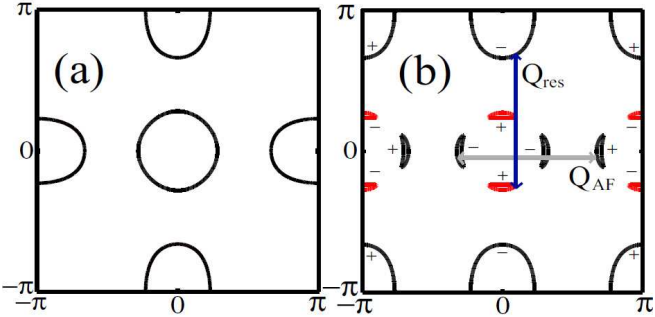}
\caption{(color online) Calculated Fermi surface topology in the normal (a) state and AF (b) state with ${\bf Q}_{1}=(\pi,0)$ ordering momentum. The $+$ and $-$ signs refer to the phase of the superconducting s$^{+-}$-wave order parameter in the coexistence phase. The arrows refer to the AF ordering wave vector and the wavevector of the incommensurate resonance excitations below T$_c$.}
\label{fig1}
\end{figure}

The electronic structure of the iron-based superconductors can be approximated by the circular hole pockets centered around the $\Gamma = (0,0)$ point of the Brillouin Zone (BZ) and
elliptic electron pockets centered around the $(\pi,0)$ and $(0,\pi)$ point of the BZ
(with 1 Fe ion per unit cell). For our purposes, it suffices to consider the simple two-band model\cite{brydon}
%
\begin{eqnarray}
\label{Ho}
\lefteqn{H  =   \sum_{\mathbf{k}, \sigma} \left\lbrace \epsilon_c(\mathbf{k}) c_{\mathbf{k} \sigma}^{\dagger} c_{\mathbf{k} \sigma}+
\epsilon_f(\mathbf{k}) f_{\mathbf{k} \sigma}^{\dagger} f_{\mathbf{k} \sigma} \right\rbrace +}&& \nonumber\\
&& \sum_{\mathbf{k},\mathbf{k'},\mathbf{q}, \sigma,\sigma'} \left\lbrace u_1 c_{\mathbf{k+q} \sigma}^{\dagger} f_{\mathbf{k'-q} \sigma'}^{\dagger} f_{\mathbf{k'} \sigma'} c_{\mathbf{k} \sigma}+ \right. \nonumber\\
&& \left.\frac{u_3}{2}\left( f_{\mathbf{k+q} \sigma}^{\dagger} f_{\mathbf{k'-q} \sigma'}^{\dagger} c_{\mathbf{k'} \sigma'} c_{\mathbf{k} \sigma} + h.c. \right)
\right\rbrace
\end{eqnarray}
where $\epsilon_c(\mathbf{k})  =   2.0  t \left( \cos{k_x}+\cos{k_y}\right) + \epsilon_c +\mu$ and
$\epsilon_f(\mathbf{k})  =   2.0 t  \left( \cos{k_x} \cdot \cos{k_y}\right) -t' \left( \cos{k_x}+\cos{k_y}\right) + \epsilon_f  +\mu$ refer to the hole ($c$) and electron ($f$) bands.
We further set $t=1.0$, $t'=0.7 t$, $\epsilon_f=1.3 t$, $\epsilon_c=-3.3 t$, and $\mu=0.03 t$ with the resulting Fermi surface shown in Fig.\ref{fig1}(a). The values of the hopping integrals slightly vary from those of Ref.\cite{brydon} to include finite doping and ellipticity. $u_1$ and $u_3$
refer to the Fermi-liquid like interactions which give rise to the AF and SC order with $s^{+-}$ symmetry.\cite{Chubukov2008}
In the following we concentrate on the coexistence phase \cite{vorontsov,fernandes,parker} in which $s^{+-}$ superconductivity coexists with antiferromagnetic order in a rather broad range of parameters.

{\em Method.} We decouple the interaction part of the Hamiltonian with respect to the AF and SC order, assuming
T$_c<$T$_{N}$. We first apply the unitary transformations with respect to the AF state and introduce the new quasiparticle operators $\alpha_{{\bf k},\sigma}$, $\beta_{{\bf k},\sigma}$ for the two resulting bands $E_{\mathbf{k}}^{\alpha,\beta} = \epsilon_{\bf k}^{+}  \pm \sqrt{\left(  \epsilon_{\bf k}^{-}\right)^2 + W^2 }$. Here, $W  =   -\frac{u_{SDW}}{2} \sum_{\mathbf{k}} \left\langle c_{\mathbf{k}\sigma}^{\dagger} f_{\mathbf{k}\sigma}\right\rangle \mbox{sqn}(\sigma)$ with $u_{SDW}=u_1 +u_3$ is the AF order parameter with ${\bf Q}_1$ ordering momentum and $\epsilon_{\bf k}^{\pm}=\frac{\epsilon_c (\mathbf{k}) \pm \epsilon_{f}(\mathbf{k})}{2}$. In the AF state with ordering at ${\bf Q}_1 = (\pi,0)$ the system remains a metal as the pocket at $(0,\pi)$ remains intact. In addition,  because of the finite doping and ellipticity of the electron pockets, tiny pockets of electron and hole character remain around the $\Gamma$ point of the BZ  at intermediate values of the AF order as shown in Fig.\ref{fig1}(b).

In the next step we rewrite the SC pairing interaction in the $s^{+-}$ channel with the unitary AF transformation, and subsequently
perform a mean-field (MF) decoupling in the particle-particle channel,
keeping only anomalous expectation values of $\left\langle \alpha^{\dagger}\alpha^{\dagger}\right\rangle $ and $\left\langle \beta^{\dagger}\beta^{\dagger}\right\rangle $. The resulting MF
Hamiltonian is then diagonalized by two independent
Bogolyubov transformations for the $\alpha$ and $\beta$ band, yielding the new energy dispersions $\Omega_{\mathbf{k}}^{\gamma} = \sqrt{\left( E_{\mathbf{k}}^{\gamma} \right)^2 + \left( \Delta_{\mathbf{k}}^{\gamma}\right)^2 }$ where $\gamma=\alpha,\beta$.
The SC gaps in the $s^{+-}$ channel, $\Delta_{\mathbf{k}}^{\gamma}$ are determined self-consistently from two coupled
gap equations
\begin{eqnarray}
\label{gaps1}
\Delta_{\mathbf{k}}^{\alpha}  & = & -u_{\mathbf{k}}^2 \Delta_1-v_{\mathbf{k}}^2 \Delta_2 \nonumber \\
\Delta_{\mathbf{k}}^{\beta}  & = &   v_{\mathbf{k}}^2 \Delta_1+u_{\mathbf{k}}^2 \Delta_2
\end{eqnarray}
where
%
$\Delta_{1}  =  -\frac{u_{sc}}{2} \sum_{\mathbf{p}} \left\lbrace  v_{\mathbf{p}}^2 \left\langle \alpha_{\mathbf{p}\uparrow}^{\dagger} \alpha_{\mathbf{-p}\downarrow}^{\dagger}\right\rangle +u_{\mathbf{p}}^2 \left\langle \beta_{\mathbf{p}\uparrow}^{\dagger} \beta_{\mathbf{-p}\downarrow}^{\dagger}\right\rangle  \right\rbrace $,
$ \Delta_{2} =  +\frac{u_{sc}}{2} \sum_{\mathbf{p}} \left\lbrace  u_{\mathbf{p}}^2 \left\langle \alpha_{\mathbf{p}\uparrow}^{\dagger} \alpha_{\mathbf{-p}\downarrow}^{\dagger}\right\rangle +v_{\mathbf{p}}^2 \left\langle \beta_{\mathbf{p}\uparrow}^{\dagger} \beta_{\mathbf{-p}\downarrow}^{\dagger}\right\rangle  \right\rbrace$. Here, $u_{\mathbf{k}}^2$, $v_{\mathbf{k}}^2 = \frac{1}{2}\left( 1 \pm \frac{\epsilon^{-}_{\bf k} }{\sqrt{\left(\epsilon^{-}_{\bf k}\right)^2 + W^2 }}\right)$, $\left\langle \gamma_{\mathbf{k}\uparrow}^{\dagger}\gamma_{\mathbf{-k}\downarrow}^{\dagger}\right\rangle = \frac{\Delta_{\mathbf{k}}^{\gamma}}{2\Omega_{\mathbf{k}}^{\gamma}} \tanh{\frac{\Omega_{\mathbf{k}}^{\gamma}}{2k_BT}}$ and $u_{sc}=u_3$. The folding of the electron and hole bands with opposite sign of the superconducting gap \cite{parker} at {\bf Q}$_{1}$ does not produce nodal lines at the intersection points.\cite{footnote}
Finally, Eqs.(\ref{gaps1}) have to be supplemented by the self-consistent equation for the AF order itself in the coexistence phase.
$1  =  -\frac{u_{SDW}}{4} \sum_{\mathbf{k}} \frac{1}{\sqrt{\left(  \epsilon^{-}_{\bf k}\right)^2 + W^2 }} \left\lbrace \frac{E_{\mathbf{k}}^{\alpha}}{\Omega_{\mathbf{k}}^{\alpha}} \left[ f(\Omega_{\mathbf{k}}^{\alpha}) -\frac{1}{2} \right] - \frac{E_{\mathbf{k}}^{\beta}}{\Omega_{\mathbf{k}}^{\beta}} \left[ f(\Omega_{\mathbf{k}}^{\beta}) -\frac{1}{2} \right] \right\rbrace$.
The use of the sequential transformations is valid when the magnitude of the SC gaps is small compared to the SDW gap. However, more generally it is possible to show that the enhanced symmetry of the problem makes this approach valid for arbitrary ratios $|\Delta_i| / W$ as long as the SC gaps of the two bands are similar.

To obtain the spin susceptibility we employ a generalized
random phase approximation approach in the multiband case\cite{brydon,Knolle2010,graser}.  In terms of Green's functions, the dynamical susceptibility tensor for the longitudinal, $zz$, and
the transverse, $+-$, components is defined as
\begin{eqnarray}
\chi_{ba}^{st,m}({\bf q},i\Omega) &  = &  -\frac{1}{2\beta} \sum_{\omega_n} \sum_{{\bf p}, \sigma}
\left[G^{bs}_{{\bf p} \sigma\sigma}(i\omega_n) G^{ta}_{{\bf p+q} \sigma'\sigma'} (i\omega_n +i\Omega) \mp \right.\nonumber\\
&& \left.F^{{\star}bs}_{{\bf p} \sigma\bar{\sigma}}(i\omega_n) F^{ta}_{{\bf p+q} \bar{\sigma}'\sigma'} (i\omega_n +i\Omega) \right]\quad,
\end{eqnarray}
where $m=zz,+-$ and $\sigma=\sigma'$ ($\sigma=\bar{\sigma}'$) for the longitudinal (transverse) component. Here,
%
$G_{{\bf p} \sigma \sigma^{\prime}}^{st}(i\omega_n)=-\int_{0}^{\beta} d \tau \langle T_{\tau}
s_{{\bf p}\sigma}(\tau)t_{{\bf p}\sigma^{\prime}}^{\dagger}(0)
\rangle e^{i w_n \tau}$, and $F_{{\bf p} \uparrow \downarrow}^{st}(i\omega_n)=-\int_{0}^{\beta} d \tau \langle T_{\tau}
s_{{\bf p}\uparrow}(\tau)t_{{\bf p}\downarrow}(0)
\rangle e^{i w_n \tau}$.
Then, the spin susceptibility is obtained via a Dyson equation
$\left[\chi_{ba}^{st}\right]_{RPA}=\chi_{ba}^{st}+\left[\chi_{b'a'}^{st}\right]_{RPA} U_{c'd'}^{b'a'}\chi_{ba}^{c'd'}$.


In the following we evaluate numerically the gap equations for the AF state with ${\bf Q}_{1}=(\pi,0)$ ordering wave vector and $s^{+-}$ SC self-consistently at T=10K.  We choose the following interaction parameters that allow for a self-consistent solution of the AF and SC gaps at this temperature: $u_{SDW}\approx 4.9$t and $u_3 \approx 5.5$t. Correspondingly we obtain a AF gap $W=0.097$t and SC gaps $\Delta_1=0.037t$, $\Delta_2=0.016t$.
We note that self-consistency is crucial to obtain the correct spectrum of the spin excitations in the AF and AF+SC state.

{\em Results.} To build up the picture for the coexistence phase,  we  consider the situation of the pure AF and $s^{+-}$ SC phases. In particular, setting SC gaps $\Delta_1=\Delta_2=0$ we find $W=0.136t$, significantly larger than the AF gap coexisting with SC because both orders compete for the same FS and suppress each other.
\begin{figure}[h]
\centering
\includegraphics[width=1.0\linewidth]{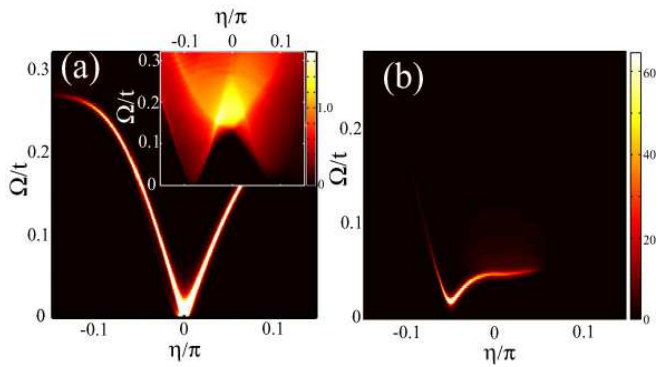}
\caption{(color online) The (${\bf q},\Omega$) intensity map (in states/t) of the imaginary part of the transverse spin susceptibility  around the AF wave vector ${\bf Q}_{1} = (\pi,0)$ along a $(\pi,0.15\pi) \to (\pi,0) \to (1.15\pi,0)$ cut of the BZ, parametrized by $\eta$, in the pure AF (a) and pure s$^{+-}$ superconducting (b) states. The inset in (a) shows the imaginary part of the transverse spin susceptibility  around ${\bf Q}_2 = (0,\pi)$. }
\label{fig2}
\end{figure}
In Fig.\ref{fig2}(a) we show the imaginary part of the transverse spin susceptibility around the magnetic ordering wave vector in the AF state only. The dispersion of the spin waves is strongly anisotropic in $q_x$ ($\eta<0$) and $q_y$ ($\eta>0$) direction due to ellipticity of the electron pockets involved, well visible up to an energy of $2W$.\cite{Knolle2010} Despite the mismatch between the sizes of electron and hole pockets separated by {\bf Q}$_{1}$, the spin waves are commensurate.

In contrast to the gapless behavior of the transverse spin excitations around the ordering vector {\bf Q}$_1$, they are gapped near ${\bf Q}_2$ as shown in the inset of Fig.\ref{fig2}. This gap, $ \Omega_2 \approx 0.1t$, is a consequence of the ellipticity of the electron pockets which lowers the symmetry of the model and favors the stripe magnetic structure over magnetic orders involving both momenta, ${\bf Q}_1$ and ${\bf Q}_2$.\cite{eremin2010,Knolle2010} For decreasing $W$, the spin excitations are still peaked around $\Omega_2$ but they are not necessarily gapless due to the onset of the particle-hole continuum, characterized by $\Omega_{ph}\sim W$. The latter originates from the scattering between tiny remnant FS pockets and the electron Fermi surface centered around $(0,\pi)$ as depicted in Fig.\ref{fig1}(b). For $\Omega_{ph} << \Omega_2$,  the excitations around {\bf Q}$_2$ are thus overdamped paramagnons. This is further supported by the fact that the value of $\Omega_2$ is the same in the longitudinal, Im$\chi^{zz}$, and the transverse, Im$\chi^{+-}$, part of the spin susceptibility. This contrasts to the behavior near {\bf Q}$_{1}$ where the spin rotational symmetry is explicitly broken, {\it i.e.} the Goldstone mode forms in Im$\chi^{+-}$ and the longitudinal excitations are gapped by twice the AF gap energy, 2$W$.

Next we set the AF gap $W=0$ and solve the mean-field gaps for the superconducting s$^{+-}$ state self-consistently obtaining $\Delta_1 \approx0.01t$ and $\Delta_2 \approx 0.04t$. We note that this limit has to be taken with care as our formalism initially relies on the assumption that $W\gg \Delta_i$ or similar magnitudes of the SC gaps on the electron and hole pockets.

In addition, the correct behavior of the superconducting gap for $W \to 0$ has to be taken. The results are shown in Fig.\ref{fig2}(b). Due to the fact that $\Delta_{\bf k} = - \Delta_{{\bf k+Q}_{AF}}$ in the s$^{+-}$ superconductor, a dispersing resonance is seen at nonzero energy below twice the SC gap values.\cite{maier,korshunov} However, in contrast to the pure AF, the position of the resonance is at the incommensurate momentum, ${\bf Q}_{res} \approx (1.05\pi,0)$. The incommensuration is induced by the specific shape of the FS with different sizes of the electron and hole pockets. \cite{Maiti}

%

\begin{figure}[t]
\centering
\includegraphics[width=1.0\linewidth]{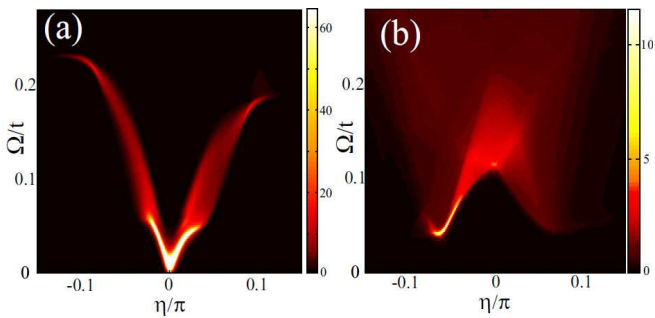}
\caption{ (color online) The (${\bf q},\Omega$) intensity map (in states/t) of the imaginary part of the transverse spin susceptibility  around the AFM wave vector ${\bf Q}_1 = (\pi,0)$ along $(\pi,0.15\pi) \to (\pi,0) \to (1.15\pi,0)$ direction of the BZ  (a) and around the ${\bf Q}_2 = (0,\pi)$ along $(0, 1.15\pi) \to (0, \pi) \to (0.15 \pi, \pi) $ direction in the coexistence SC+AF phase.}
\label{fig3}
\end{figure}

Finally, in Fig.\ref{fig3} we present the results for coexisting AF+SC orders. In Fig.\ref{fig3}(a) we show the evolution of the transverse spin excitations around {\bf Q}$_1$. Due to the finite magnetization, well-defined anisotropic spin waves exist around the ordering momentum similar to the pure AF state.  This is expected as the Goldstone mode in the AF phase reflects the breaking of the spin-rotational symmetry of the system and as such is not affected by the presence of the additional spin-singlet superconducting state. At the same time, and in contrast to the pure AF state, spin waves are damped away from ${\bf Q}_1$ at energies well below twice the AF gap 2$W$. In particular, additional damping starts at energies of twice the superconducting gap.  This new feature of the coexistence phase arises due to the renormalization of the particle-hole (p-h) continuum. In the AF, the p-h response is determined by the transition between $\alpha-$ and $\beta-$bands as well as the intraband transitions. The latter are suppressed by the vanishing AF matrix elements around the ordering wave vector. This also guarantees the stability of the Goldstone mode even in the situation when the AF gap does not gap the entire FS and remnant electron and hole pockets are still present. The damping of the spin waves arises from interband transitions. They are gapped exactly by $2W$  at {\bf Q}$_{1}$; however, this gap becomes progressively smaller away from it. In the coexistence phase, the interband excitations at {\bf Q}$_{1}$ are gapped by the sum of twice SC and AF gaps. Away from {\bf Q}$_{1}$ the effect of the AF quickly vanishes.
However, the p-h continuum is still gapped  by approximately twice the SC gap. Above this energy it experiences a discontinuous jump due to $\mbox{sgn}(\Delta_{\bf k}^{\alpha}) = - \mbox{sgn}(\Delta_{\bf k}^{\beta})$. As a result, the spin wave excitations are damped at energies associated with twice the SC gap as seen from Fig.\ref{fig3}(a). Note that the 'stronger' damping of the spin waves in the coexistence phase [Fig.\ref{fig3}(a)] originates from the smaller value of the AF gap in the coexistence phase. The discontinuous jump of the particle-hole continuum at approximately $2\Delta_0$ is absent in the pure AF state.

In Fig.\ref{fig3}(b) we show the spin excitations in the coexistence phase around ${\bf Q}_2$. As discussed above, in the pure AF phase the excitations around {\bf Q}$_2$ are paramagnons centered at frequency $\Omega_2$ and almost isotropic in spin space. In the coexistence phase, however, the SC gap modifies the behavior of the p-h continuum around {\bf Q}$_2$ which was gapless in the AF only case when tiny hole pockets centered around $(0,0)$ were still present. The SC gaps the p-h continuum up to $2\Delta_0$. In addition, due to the change of sign of the gaps between the pockets, the p-h continuum experiences a discontinuous jump at this frequency and the necessary condition for the formation of the excitonic resonance at energies below 2$\Delta_0$ forms. It is remarkable that this resonance exist at energies which are much lower than the energy position of the original mode at $\Omega_2$ in the AF state.

The incommensuration of the resonance in the coexistence phase is closely related to the reconstructed FS for which hole pocket states that would be connected  by ${\bf Q_2}$ to an electron pocket are gapped for sufficiently large magnetic order. Note also that contrary to the ordering momentum, the excitations around ${\bf Q}_2$ remain isotropic in spin space, {\it i.e.} the excitations in the longitudinal and transverse channel show qualitatively similar behavior.

Overall the results in the coexistence region show the peculiarity of the spin excitations in the ferropnictides:
while around the ordering momentum ${\bf Q}_1$ the excitations are determined by the presence of the AF order, the excitations around ${\bf Q}_2$  are dominated by the SC and the formation of the incommensurate spin resonance. The formation of the resonance  effectively lowers the energy splitting between the different magnetic states in the coexistence region. Correspondingly the system tends to increase its magnetic degeneracy in the presence of $s^{+-}$ superconductivity. As a consequence, the collective excitations associated with the presence of the incommensurate spin resonance close to ${\bf Q}_2$ may occur in the coexistence region, a phenomenon which would be absent in the pure AF state. This possibility would be interesting to check experimentally.

A direct probe to see the excitations around ${\bf Q}_1$ and ${\bf Q}_2$ separately in INS experiments is to look at untwinned crystals. However, existing experiments on twinned crystals already indicate that both spin wave and spin resonance are present in the coexistence phase. Ref.\cite{pratt} found that spin excitations in the coexistence AF+SC phase show strong anisotropy along the $c$-crystallographic direction, absent in the pure SC state. The difference in energy of the spin excitations at $(\pi,\pi,q_z)$ is $3-4$meV between $q_z = 0$ and $q_z = \pi$. At the same time, it was argued\cite{park} that due to the peculiar crystallographic structure
of 122 compounds, the excitations at $(\pi,\pi,0)$ and $(\pi,\pi,\pi)$ in the folded BZ zone correspond to  excitations at $(\pi,0,0)$ and $(0,\pi,\pi)$ wave vectors in the unfolded BZ, respectively. In other words, while excitations around $(\pi,\pi,0)$ refer to the true spin waves at {\bf Q}$_1$ (gapped in the experimental system by
spin-orbit coupling we have omitted) the excitations around $(\pi,\pi,\pi)$ originate from the incommensurate resonance at ${\bf Q}_2$. The crucial test to see whether this is true is to study the spin rotational symmetry of these excitations and their possible incommensurability.

We have analyzed spin excitations in the coexistence phase of iron-based superconductors in which both AF and s$^{+-}$ SC coexist as a model of an approximate emergent symmetry in an itinerant magnet.  We find that the excitations around the 'failed' ordering wavevector ${\bf Q}_2$  become resonant at
energies $\Omega_{res} < 2\Delta_0$.
In addition, the longitudinal component of the spin susceptibility differs strongly near ${\bf Q}_1$ and ${\bf Q}_2$. Around ${\bf Q}_1$, where the transverse excitations are gapless Goldstone modes, longitudinal excitations are gapped by 2W on account of the presence of AF order and are only weakly affected by the SC; while at ${\bf Q}_2$, excitations are isotropic with a gap now set by the superconductivity, 2$\Delta_0$.

We thank A. Chubukov, R. Fernandes, P. Hirschfeld, A. Goldman, S. Maiti, R. McQueneey, and W. Wang for useful discussions.
 IE acknowledges the financial support of the SFB Transregio 12 and DAAD (PPP USA No. 50750339). JK acknowledges
 support from a Ph.D. scholarship from the Studienstiftung
 des deutschen Volkes and the IMPRS Dynamical Processes
 in Atoms, Molecules and Solids. IE and JK were supported by ICAM travel grants.


\begin{thebibliography}{99}

\bibitem{kamihara} Y. Kamihara, T. Watanabe, M.  Hirano and  H. Hosono, J. Am. Chem. Soc. {\bf 130}, 3296 (2008).

\bibitem{podolsky} D. Podolsky, H.-Y. Kee, and Y.B. Kim, Europhys. Lett. {\bf 88}, 17004 (2009).

\bibitem{Chubukov2008} A. V. Chubukov, D. V. Efremov, and I. Eremin, Phys. Rev. B {\bf78} 134512 (2008).


\bibitem{chandra} P. Chandra, P. Coleman, and A. I. Larkin, Phys. Rev. Lett. {\bf 64}, 88 (1990).



\bibitem{eremin2010} I. Eremin and A.V. Chubukov, Phys. Rev. B {\bf 81}, 024511 (2010).

\bibitem{Knolle2010} J. Knolle, I. Eremin, A.V. Chubukov, and R. Moessner, Phys. Rev. B {\bf 81}, 140506(R) (2010).

\bibitem{yaresko} A. N. Yaresko, G.-Q. Liu, V. N. Antonov, and O. K. Andersen, Phys. Rev. B {\bf 79}, 144421 (2009).

\bibitem{maier} T.A. Maier and D. J. Scalapino, Phys. Rev. B {\bf 78},
020514(R) (2008).

\bibitem{korshunov} M. M. Korshunov and I. Eremin, Phys. Rev. B {\bf 78},
140509(R) (2008).

\bibitem{brydon} P.M.R. Brydon, and C. Timm, Phys. Rev. B {\bf 80}, 174401 (2009).

\bibitem{vorontsov} A.B.Vorontsov, M.G.Vavilov, A.V.Chubukov, Phys. Rev. B {\bf 81} , 174538 (2010).

\bibitem{fernandes} R. M. Fernandes, J\"org Schmalian, Phys. Rev. B {\bf 82}, 014521 (2010).

\bibitem{parker} D. Parker, M.G. Vavilov, A.V. Chubukov, I.I. Mazin, Phys. Rev. B {\bf 80}, 100508(R) (2009).

\bibitem{footnote} In contrast to Ref.\cite{parker}, in our choice of gauge the remnant pockets around the $\Gamma$ point have opposite signs, see Fig\ref{fig1}(b).

\bibitem{graser} S. Graser, T. A. Maier, P. J. Hirschfeld, and D. J. Scalapino, New
J. Phys. {\bf 11}, 025016 (2009).

\bibitem{Maiti} S. Maiti, J. Knolle, I. Eremin, A. V. Chubukov, arXiv:1108.0266 (unpublished).




\bibitem{pratt} D.K. Pratt {\it et al.}, Phys. Rev. B {\bf 81}, 140510(R) (2010).



\bibitem{park} J.T. Park {\it et al.}, Phys. Rev. B {\bf 82}, 134503 (2010).





%
%

\end{thebibliography}
\end{document}